\documentclass{llncs}
\usepackage{amsmath}
\usepackage{amssymb}
\usepackage{tikz}
\usepackage{mathtools}
\usepackage{wrapfig}

\DeclarePairedDelimiterX{\infdivx}[2]{(}{)}{%
	#1\;\delimsize\|\;#2%
}

\DeclarePairedDelimiter\abs{\lvert}{\rvert}

\title{A Novel Method for Epileptic Seizure Detection Using Coupled Hidden Markov Models}

\author{Jeff Craley\inst{1} \and Emily Johnson\inst{2} \and Archana Venkataraman\inst{1}}
\institute{Dept. of Electrical and Computer Engineering Johns Hopkins University \and Department of Neurology, Johns Hopkins Medical Institute}

\begin{document}

\maketitle

\begin{abstract}
We propose a novel Coupled Hidden Markov Model to detect epileptic seizures in multichannel electroencephalography (EEG) data. Our model defines a network of seizure propagation paths to capture both the temporal and spatial evolution of epileptic activity. To address the intractability introduced by the coupled interactions, we derive a variational inference procedure to efficiently infer the seizure evolution from spectral patterns in the EEG data. We validate our model on EEG aquired under clinical conditions in the Epilepsy Monitoring Unit of the Johns Hopkins Hospital. Using 5-fold cross validation, we demonstrate that our model outperforms three baseline approaches which rely on a classical detection framework. Our model also demonstrates the potential to localize seizure onset zones in focal epilepsy.
\end{abstract}

\section{Introduction}

Epilepsy is a heterogenous neurological disorder characterized by recurring and unprovoked seizures \cite{miller2014}. 
It is estimated that 20-40\% of epilepsy patients are medically refractory and do not respond to drug therapy. 
Alternative therapies for these patients crucially depend on being able to detect epileptic activity in the brain.
The most common modality used for seizure detection is multichannel electroencephalography (EEG) acquired on the scalp.
The clinical standard for seizure detection involves visual inspection of the EEG data, which is time consuming and requires extensive training.
In this work, we develop an automated seizure detection procedure for clinically acquired multichannel EEG recordings.

There is a vast body of literature on epileptic seizure detection from a variety of viewpoints. The nonlinearity of EEG signals has inspired the application of  techniques from chaos theory such as approximate entropy and Lyapunov exponents as in \cite{acharya2012automated} and \cite{guler2005recurrent}, respectively.
Alternatively, wavelet and other time-frequency based features seek to capture the non-stationarity of the EEG signal as in \cite{zandi2010automated}.
These features are fed into standard classification algorithms to detect seizure activity.
A fundamental limitation of the methods in \cite{acharya2012automated} and \cite{guler2005recurrent} is that they are trained on a single channel of EEG data and fail to generalize in practice. Multichannel strategies such as those in \cite{zandi2010automated,hunyadi2012incorporating,shoeb2010application,baldassano2016novel} rely heavily on prior seizure recordings to train patient specific detectors, which are often unavailable.

Unlike prior work, our approach explicitly models the spatial dynamics of a seizure through the brain over time. We build on existing work in Hidden Markov Models (HMMs) \cite{baldassano2016novel}, adopting a Coupled HMM (CHMM) \cite{brand1997coupled} to model interchannel dependencies. Specifically, the likelihood that an EEG channel will transition into a seizure state will increase if neighboring channels are in a seizure state. This coupling renders exact inference intractable. Therefore we develop a variational Expectation Maximization (EM) algorithm for our framework.

We evaluate our algorithm using 90 scalp EEG recordings from 15 epilepsy patients acquired in the Epilepsy Monitoring Unit (EMU) of the Johns Hopkins Hospital. These recordings contain up to 10 minutes of baseline activity before and after a seizure and have not been screened for artifacts. 
We compare our CHMM to classifiers evaluated on a framewise and channelwise basis. Our algorithm outperforms these baselines and demonstrates efficacy in classifying seizure intervals. Our algorithm provides localization information that could be useful for determining the seizure onset location in cases of focal epilepsy.

\section{Generative Model of Seizure Propagation}
We adopt a Bayesian framework for seizure detection. The latent variables $\mathbf{X}$ denote the seizure or non-seizure states. $\mathbf{Y}$ corresponds to observed data feature vectors computed from EEG channels as shown in Fig. \ref{fig:model} (a). The random variable $X^t_i$ denotes the latent state of EEG channel~$i$ at time $t$. We assume three possible states: pre-seizure baseline ($X^t_i = 0$), seizure propagation ($X^t_i = 1$), and post-seizure baseline ($X^t_i = 2$). The corresponding observed ``emission'' feature vectors $Y^t_i$ are continuous statistics computed from time window~$t$ of the EEG channel~$i$. For convenience, we also define the ensemble variables $\mathbf{X}^t \triangleq \left[ X^t_1, \dots, X^t_N \right]^T$ where $N$ is the number of electrodes. Given the EEG observations, our goal is to infer the latent seizure state for each chain at all times.

\subsection{Model Formulation and Inference}
\label{sub:model}
\begin{figure}[h]
\centering
\begin{minipage}{0.54\textwidth}
	\centering
	\centerline{\includegraphics[width=\textwidth]{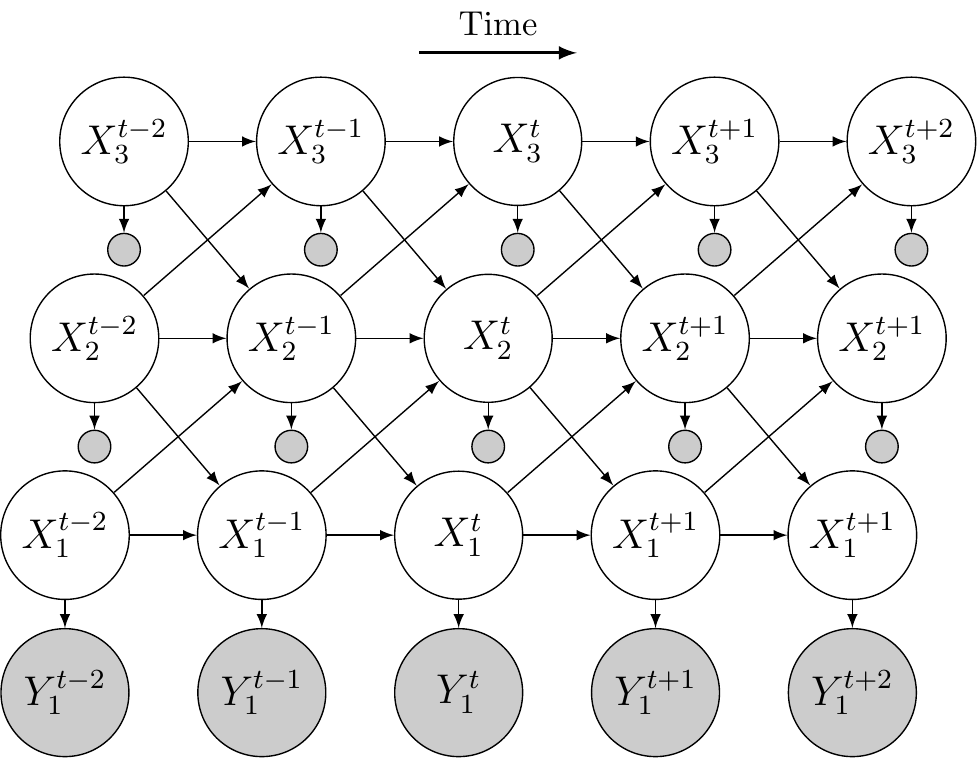}}
	\centerline{\footnotesize (a)}
\end{minipage}
\begin{minipage}{0.40\textwidth}
	\centering
	\centerline{\includegraphics[width=\textwidth]{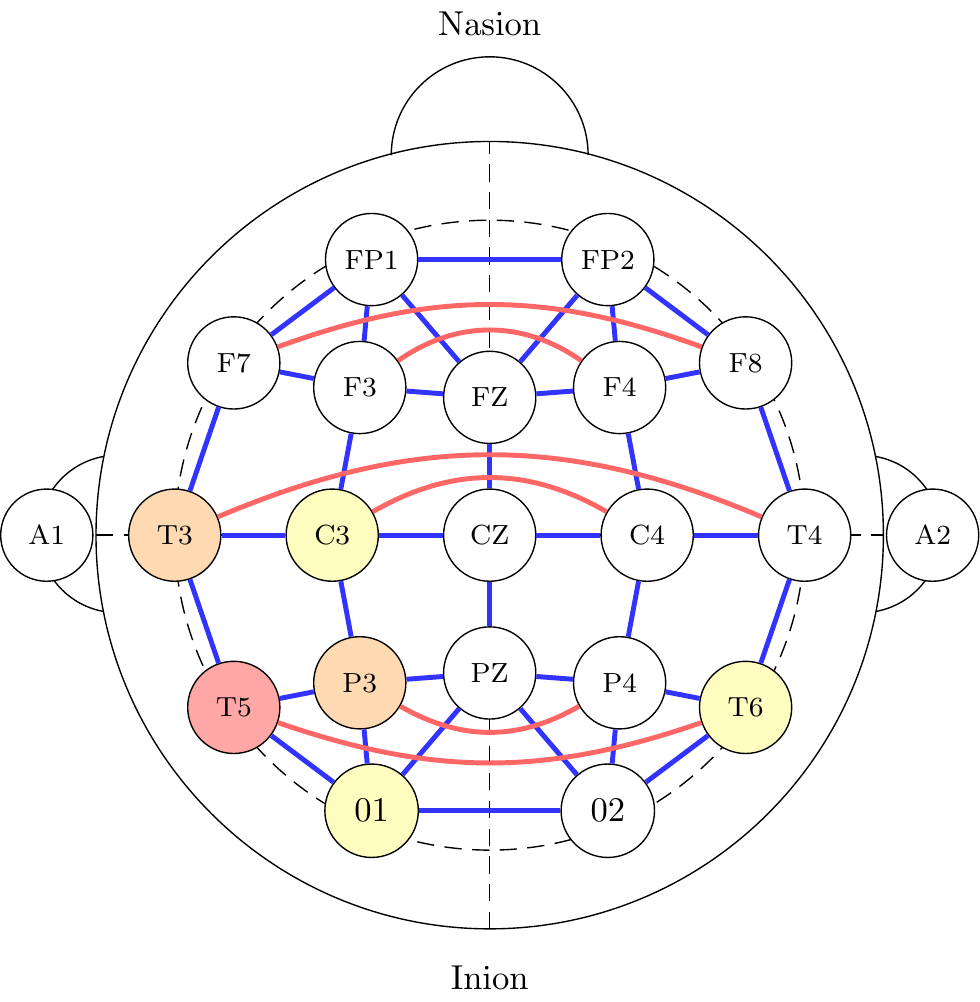}}
	\centerline{(b)}
\end{minipage}
\caption{(a) Graphical model depicting a three chain CHMM. Observed nodes are shaded gray, while latent nodes are shown in white. (b) EEG channels in the 10/20 international system \cite{jurcak200710}, cross hemispheric (red) and neighboring (blue) channel connections. A seizure propagates from the red, to the orange, and finally yellow shaded electrodes.}
\label{fig:model}
\end{figure}

Fig. \ref{fig:model} (b) shows the coupling between electrodes of the 10/20 international system \cite{jurcak200710}. We define the aunts $au(\cdot)$ of a given node as the set of electrodes connected to it in Fig. \ref{fig:model} (b). The joint distribution of $\mathbf{X}$ and $\mathbf{Y}$ factorizes into transition priors that depend on both a channel's own previous state and those of its aunts $P(X^t_i \mid \mathbf{X}^{t-1}_{au(i) \bigcup i})$,  and emission likelihoods $P(Y^t_i \mid X^t_i)$ as in Eq.~(\ref{eq:likelihood_factorization}). For simplicity we assume that all recordings begin in a non-seizure state $(X^0_i=0 \, \forall \, i)$.
\begin{equation}
	P(\mathbf{X},\mathbf{Y}) = \prod_{i=1}^{N} P(Y^0_i \mid X^0_i ) \prod_{t=1}^T P(Y^t_i \mid X^t_i ) P(X^t_i \mid \mathbf{X}^{t-1}_{au(i) \bigcup i} )
	\label{eq:likelihood_factorization}
\end{equation}
Note that the observed emissions are conditionally independent given the latent states. Below, we detail the model formulation and inference algorithm.

\subsubsection{Coupled State Transitions.}
\label{par:coupled}

The distribution over state vectors $\mathbf{X}^t$ forms a first order Markov chain. This distribution further factorizes into products of transition distributions of individual chains $P(\mathbf{X}^t \mid \mathbf{X}^{t-1} ) = \prod_{i=1}^N P(X^t_i \mid \mathbf{X}^{t-1}_{au(i) \bigcup i } )$. We encode these chainwise transition probabilities using time inhomogenous transition matrices as shown in Eq. (\ref{eq:transition}). This structure ensures each channel begins in a non-seizure baseline state, transitions into an active seizure state, and transitions into a post-seizure state.
\begin{equation}
	A^t_i = \left[ \begin{array}{ccc}
		1 - g^t_i	& 	g^t_i		& 0 \\
		0			& 	1 - h^t_i	& h^t_i \\
		0			&	0			& 1
	\end{array} \right]
	\label{eq:transition}
\end{equation}

The transition matrix $A^t_i$ is governed by neighboring and contralateral EEG channels to capture the main modes of seizure propagation, as shown in Fig.~\ref{fig:model}~(b). Let $\eta^t_i$ be the number of aunts in the seizure state in the previous timestep. We model the transition probabilities into and out of the seizure state via logistic regression functions of $\eta^t_i$ as shown in Eq. (\ref{eq:logistic_transitions}). Parameters $\lbrace \rho_0, \phi_0 \rbrace$ control the base onset and offset rates while $\lbrace \rho_1, \phi_1 \rbrace$ control the effects of a channel's aunts.
\begin{equation}
	\log\left(\frac{g^t_i}{1 - g^t_i}\right) = \rho_0 + \rho_1 \eta^t_i, \qquad
	\log\left(\frac{h^t_i}{1 - h^t_i}\right) = \phi_0 + \phi_1 \eta^t_i
	\label{eq:logistic_transitions}
\end{equation}

\subsubsection{Emission Likelihood.}
\label{par:emission}
We use a Gaussian Mixture Model (GMM) to describe the emissions $Y^t_i$ of each chain. Let $C^t_i$ be the mixture from which $Y^t_i$ was generated. Let $\pi^k_{ij}$ be the prior probability of mixture component $j$ when $X^t_i = k$ for $k=0,1,2$. The joint distribution over $Y^t_i$ and $C^t_i$ can be expressed as follows
\begin{equation}
	\begin{aligned}
	P(Y^t_i, C^t_i=j \mid X^t_i=k) & = P(Y^t_i \mid C^t_i=j) P(C^t_i=j \mid X^t_i=k) \\
		& = \pi^k_{ij} \mathcal{N} \left( Y^t_i; \mu_{ij}, \Sigma_{ij} \right)
	\end{aligned}
	\label{eq:emission}
\end{equation}
Effectively, the emission distributions for all observed variables share the same mean parameters $\mu_{ij}$ and covariance parameters $\Sigma_{ij}$, but use different mixture weights based on the latent seizure state $k$. We tie weights for both pre- and post-seizure states, i.e. $\pi_{ij}^0 = \pi_{ij}^2$ for all channels $i$ and mixture components $j$. The data likelihood $P(Y^t_i \mid X^t_i)$ can be computed by marginalizing over $j$.

\subsubsection{Approximate Inference Using Variational EM.}
\label{par:EM}
Exact inference for the CHMM is intractable due to the coupled state transitions. Therefore we develop a structured variational algorithm \cite{murphy2012machine}, in which we approximate the posterior distribution over $\mathbf{X}$ as a set of $N$ independent HMM chains:
\begin{equation}
	Q(\mathbf{X}) = \prod_{i=1}^N \frac{1}{Z_{Q_i}} Q_i (\mathbf{X}_i) 
		= \prod_{i=1}^N \frac{1}{Z_{Q_i}} \prod_{t=1}^T T^t_i(X^t_i \mid X^{t-1}_i) E^t_i(X^t_i) \enspace .
	\label{eq:approximating_distribution}
\end{equation}
As seen in Eq. (\ref{eq:approximating_distribution}), each approximating chain includes a normalizing constant $Z_{Q_i}$, a transition term $T^t_i(X^t_i \mid X^{t-1}_i)$, and an emission term $E^t_i(X^t_i)$. 

The transition distribution $T^t_i(X^t_i \mid X^{t-1}_i)$ is encoded by a state transition matrix $\tilde{A}^t_i$ which mimics the structure of Eq. (\ref{eq:transition}). Here $\tilde{g}^t_i$ and $\tilde{h}^t_i$ are variational transition parameters analagous to the original transition parameters $g^t_i$ and $h^t_i$.
\begin{equation}
	\tilde{A}^t_i = \left[ \begin{array}{ccc}
		1 - \tilde{g}^t_i	& \tilde{g}^t_i			& 0 \\
		0					& 1 - \tilde{h}^t_i		& \tilde{h}^t_i \\
		0					& 0						& 1
	\end{array} \right]
	\label{eq:var_transition}
\end{equation}
In contrast to Eq. (\ref{eq:var_transition}), the emission distribution $E^t_i(X^t_i)$ weighs the contribution of the observed data $Y^t_i$ through variational parameters $\tilde{l}^t_{i0}$ and $\tilde{l}^t_{i1}$. Thus $E^t_i(X^t_i=0, 2) = \tilde{l}^t_{i0}$ and $E^t_i(X^t_i = 1) = \tilde{l}^t_{i1}$.

We learn variational parameters for each chain by minimizing the free energy of the approximation. We perform this minimization by decoupling the free energy into expectations over a single channel and expectations over the remaining channels. The index ``$-i$'' in Eq. (\ref{eq:free_energy_chains}) denotes the set of channels excluding $i$.
\begin{equation}
\begin{aligned}
\mathcal{FE} & = -E_Q \left[ \log p(\mathbf{X}, \mathbf{Y}) \right] + E_Q \left[ \log Q(\mathbf{X}) \right] \\
	& = -E_{Q_i} \left[ E_{Q_{-i}} \left[ \log p(\mathbf{X}_i, \mathbf{Y}_i \mid \mathbf{X}_{-i}, \mathbf{Y}_{-i}) \right] \right] + E_{Q_{i}} \left[ \log Q_{i}({X}_{i}) \right] \\
	& \quad - E_{Q_{-i}} \left[ \log p(\mathbf{X}_{-i}, \mathbf{Y}_{-i}) \right] + E_{Q_{-i}} \left[ \log Q_{-i}({X}_{-i}) \right] \\
\end{aligned}
\label{eq:free_energy_chains}
\end{equation}
Notice that the last line of Eq. (\ref{eq:free_energy_chains}) does not depend on the parameters of chain~$i$, allowing a natural fixed point iteration over the parameters of a single chain while holding all other chains constant. This minimization fixes the variational parameters $\tilde{l}^t_i$ equal to the GMM likelihood of the observed data:
\begin{equation}
	\tilde{l}^t_{i0} = p(Y^t_i \mid X^t_{i} = 0,2), \qquad \tilde{l}^t_{i1} = p(Y^t_i \mid X^t_{i} = 1) \enspace .
\end{equation}
Similarly the updates for the variational transition parameters form logistic regressions where the activations are the expected value of the original activations.
\begin{equation}
	\log\left(\frac{\tilde{g}^t_i}{1 - \tilde{g}^t_i}\right) = \rho_0 + \rho_1 E_{Q_{au(i)}} \left[ \eta^t_i \right], \quad
	\log\left(\frac{\tilde{h}^t_i}{1 - \tilde{h}^t_i}\right) = \phi_0 + \phi_1 E_{Q_{au(i)}} \left[ \eta^t_i \right]
\end{equation}

Once the variational parameters have been computed, the approximating distribution takes the form of an HMM where the $\tilde{l}$ parameters capture the likelihood of the data under each latent state. We can use the forward-backward algorithm \cite{murphy2012machine} to compute the expected latent states $E_Q[X^t_i]$, the expected state transitions, and the expected number of aunts in the seizure state $E_Q[\eta^t_i]$.

\subsubsection{Learning the Model Parameters.}
We use the expected values of the latent states and mixture components to update the transition parameters $\lbrace \rho_{i}, \phi_{i} \rbrace$ and the emission parameters $\lbrace \mu_{ij}, \Sigma_{ij}, \pi^k_{ij} \rbrace$. Let $\tau^t_i(j, k)$ be the expectation that channel $i$  at time $t$ is in state $k$ with mixture $j$, we can update the emission parameters according to the soft counts of the occurrence of each mixture.
\begin{equation}
	\mu_{ij} = \frac{\sum_{k=0}^2 \sum_{t=0}^T \tau^t_i(j, k) Y^t_i}{\sum_{k=0}^2 \sum_{t=0}^T \tau^t_i(j, k)}, \quad
	\Sigma_{ij} = \frac{\sum_{k=0}^2 \sum_{t=0}^T \tau^t_i(j, k) \left( Y^t_i - \mu_{ij} \right)^2 }{\sum_{k=0}^2 \sum_{t=0}^T \tau^t_i(j, k)}
\end{equation}
\begin{equation}
	\pi_{ij}^0 = \pi_{ij}^2 = \frac{\sum_{t=0}^T \tau^t_i(j, 0) + \tau^t_i(j, 2)}{\sum_{j'} \sum_{t=0}^T \tau^t_i(j', 0) + \tau^t_i(j', 2)}, \quad \pi_{ij}^1 = \frac{\sum_{t=0}^T \tau^t_i(j, 1)}{\sum_{j'} \sum_{t=0}^T \tau^t_i(j', 1)}
\end{equation}

The update for the transition parameters $\lbrace \rho_i, \phi_i \rbrace$ takes the form of a weighted logistic regression. We regress the expected $\eta^t_i$ onto the expected transitions for each chain and use Newton's method to find the optimal transition parameters.

\subsubsection{Implementation Details.}
We initialize our model by training the GMM emission distributions based on the expert annotations of seizure intervals. 3 emission mixtures resulted in a reasonable compromise between sensitivity and specificity. 
Transition parameters $\rho_0$, $\rho_1$, $\phi_0$, and $\phi_1$ were initialized to -7, 2, -3, and 0, respectively. This corresponds to expected seizures every 13 minutes lasting 15 seconds, channels turning on with a 7 fold increase per aunt node, and no cross channel influence for offset.
Our EM proceedure is performed in an unsupervised fashion without further use of the labels. Model parameters are updated during the M-step of the EM algorithm. During inference, channels are updated sequentially until the scaled difference in $\mathcal{FE}$ converges to less than $10^{-4}$. 

\subsection{Baseline Comparison}
We compare our model to three alternative classification schemes. The first approach is to train a logistic regression function to distinguish between baseline and seizure intervals based on a linear combination of the EEG features. The second approach uses kernel support vector machines (SVMs) to learn a possibly nonlinear decision boundary in the EEG feature space that maximally separates the baseline and seizure conditions. Here we rely on a polynomial kernel. SVM classifiers have been used extensively for seizure detection \cite{acharya2012automated,hunyadi2012incorporating,shoeb2010application}. Finally we consider a GMM hypothesis testing scenario. This method trains GMMs for seizure and non-seizure states and classifies based on the ratio of the likelihoods under each GMM, roughly equating to our model with no transition prior.

\section{Experimental Results}
\subsection{Data and Preprocessing}
Our EEG data was recorded as part of routine clinical evaluation in the EMU of the Johns Hopkins Hospital. Our dataset consists of 90 seizures recordings from 15 patients with as much as 10 minutes of baseline before and after the seizure. Recordings were sampled at 200 Hz. We rely on expert clinical annotations denoting the seizure onset and offset to validate the performance of each method.

For preprocessing, each EEG channel was bandpass filtered through sequential application of fourth order Butterworth high and low pass filters at 1.6 Hz and 50 Hz respectively. This filtering mirrored clinical preprocessing practice for removing DC trends and high frequency components with no clinical relevance. In addition, a second order notch filter with $Q=20$ was applied at 60 Hz to the EEG recordings to remove any remaining effect of the power supply.

We considered two emission features for analysis computed from channels in common reference: the sum of spectral coefficients in brain wave frequency bands and the log line length. Features were computed on windows of 1 s with 250 ms overlap. For spectral features, a short time Fourier transform was taken after the application of a Tukey window with shape parameter 0.25. The magnitudes of the STFT coefficients corresponding to frequencies in the theta (1-4 Hz), delta (4-8 Hz), alpha (8-13 Hz), and beta (13-30 Hz) bands were summed and the logarithm was taken, resulting in a length four feature vector. The log line length was computed as the logarithm of the sum of the absolute difference between successive samples i.e. given a signal $s$ of length $T$, $\log L = \log\left( \sum_{i=0}^{T-1} \abs{s(i+1) - s(i)} \right)$.

\subsection{Seizure Detection Performance}
We use a 5 fold cross validation strategy for evaluation. Four folds were used to train each model and detection was evaluated on the held-out fold. Each recording was randomly assigned to a fold independently of patient. For our model, the training phase was used to learn the emission and transition parameters. Table \ref{tab:results} summarizes the performance for each classifier based on the average accuracy of the testing fold. The sensitivity (TPR) and specificity (TNR) denote the prediction accuracy for seizure and non-seizure frames, respectively, computed across all channels. For the probabilistic classifiers (i.e. logistic regression, GMM, and CHMM), these rates are weighted by the posterior confidence of the classifier.

\begin{figure}[htb]
\hfill
\begin{minipage}[b]{0.45\linewidth}
  \centering
  \centerline{\includegraphics[width=\textwidth]{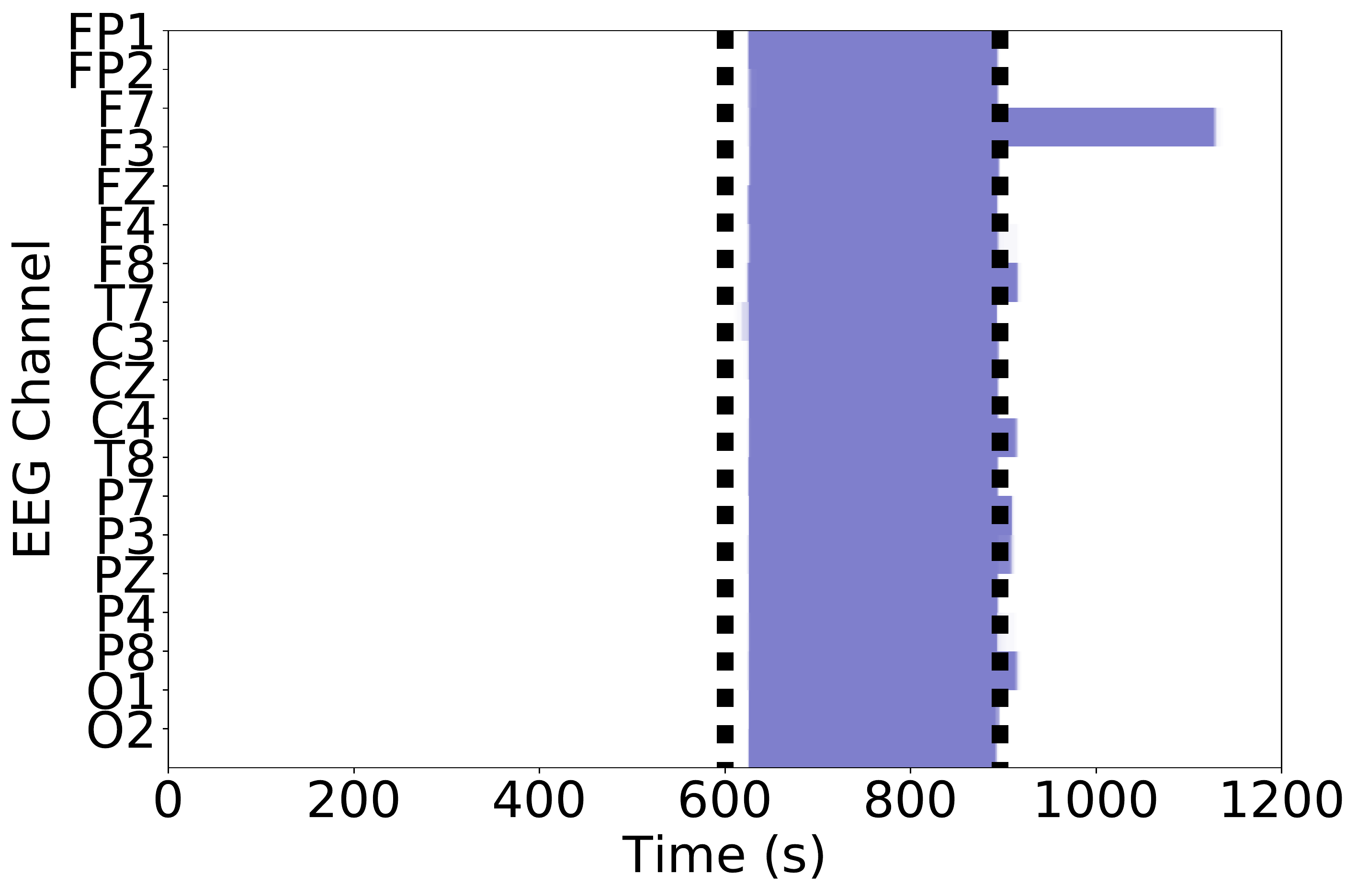}}
  \centerline{\footnotesize (a) Patient 1}
\end{minipage}
\hfill
\begin{minipage}[b]{.45\linewidth}
  \centering
  \centerline{\includegraphics[width=\textwidth]{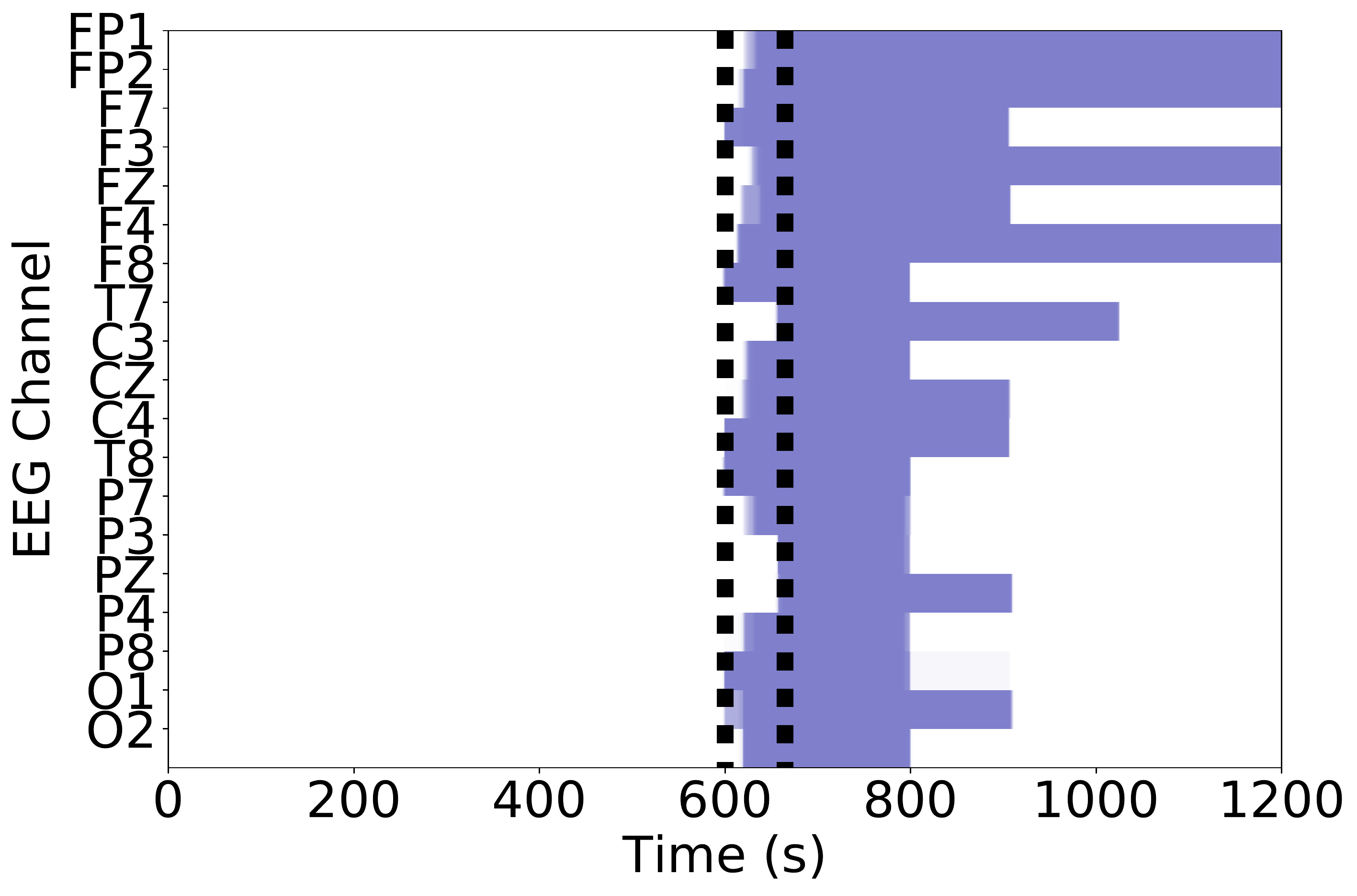}}
  \centerline{\footnotesize (b) Patient 2}
\end{minipage}
\hfill

\begin{minipage}{0.05\linewidth}

\end{minipage}
\hfill
\begin{minipage}[b]{0.90\linewidth}
  \centering
  \centerline{\includegraphics[width=\textwidth]{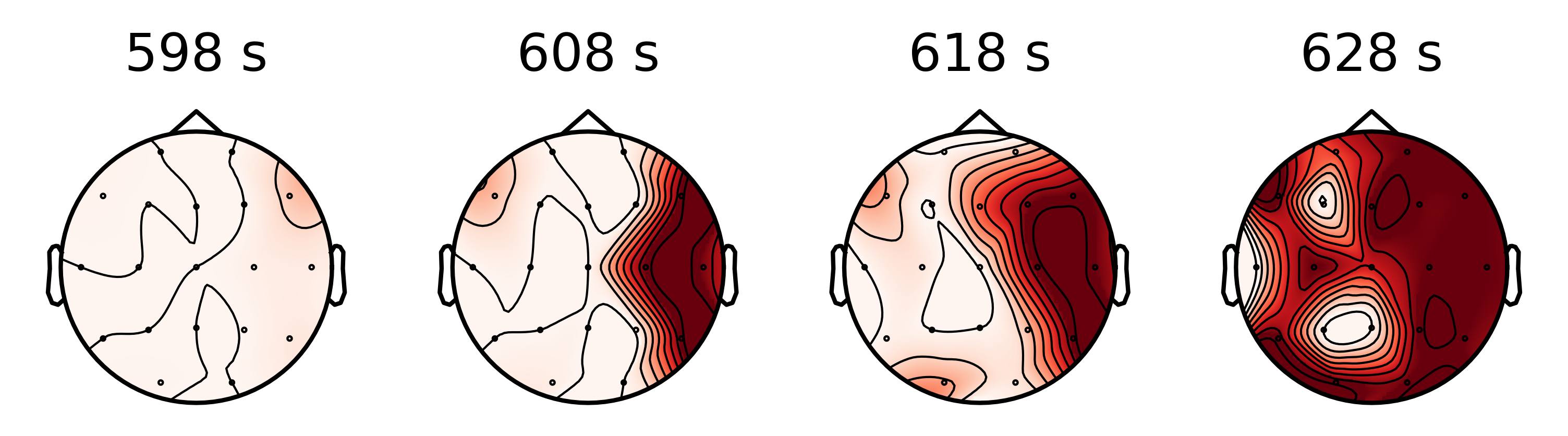}}
  \centerline{\footnotesize (c) Time lapse detail of inferred seizure onset in Patient 2}
\end{minipage}
\hfill
\begin{minipage}{0.05\linewidth}

\end{minipage}

\caption{Detection results for the CHMM. \textbf{Top row:} Estimated posterior probability of the latent seizure state for two epilepsy patients. White corresponds to pre- and post-seizure baseline while violet indicates seizure states. EEG channels corresponding to 10/20 system \cite{jurcak200710} channel locations are on the $y$ axis. The expert identified seizure region is denoted by the dashed black lines. (a) shows the models ability to accurately classify seizures across the whole brain and (b) shows the outward spread of a right temporal lobe seizure. \textbf{Bottom row:} temporal evolution of the seizure depicted in (b).}
\label{fig:inference}
\end{figure}

\begin{figure}[htb]
\begin{minipage}[b]{0.305\linewidth}
  \centering
  \centerline{\includegraphics[width=\textwidth]{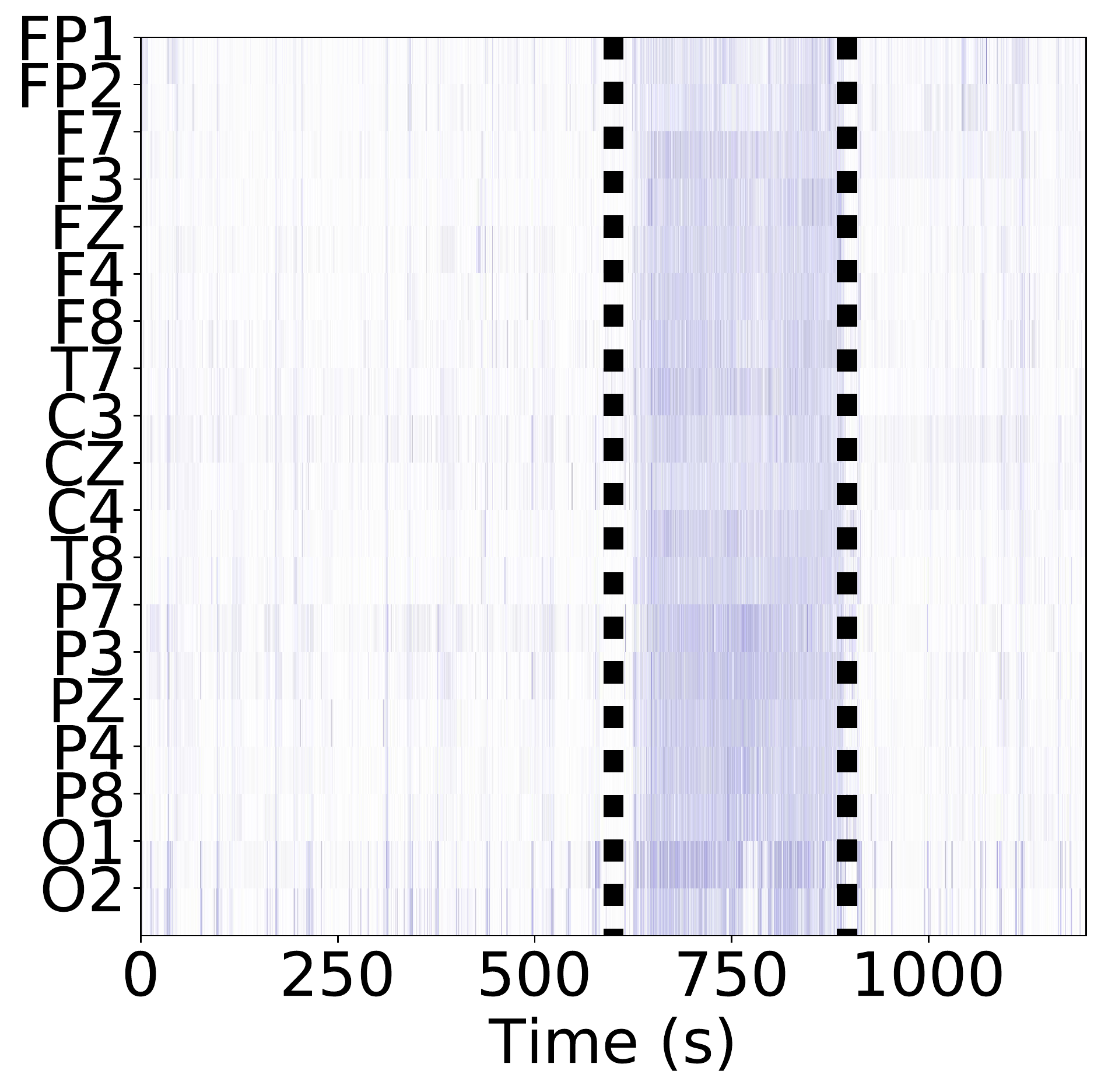}}
  \centerline{\footnotesize (a) GMM}
\end{minipage}
\hfill
\begin{minipage}[b]{.305\linewidth}
  \centering
  \centerline{\includegraphics[width=\textwidth]{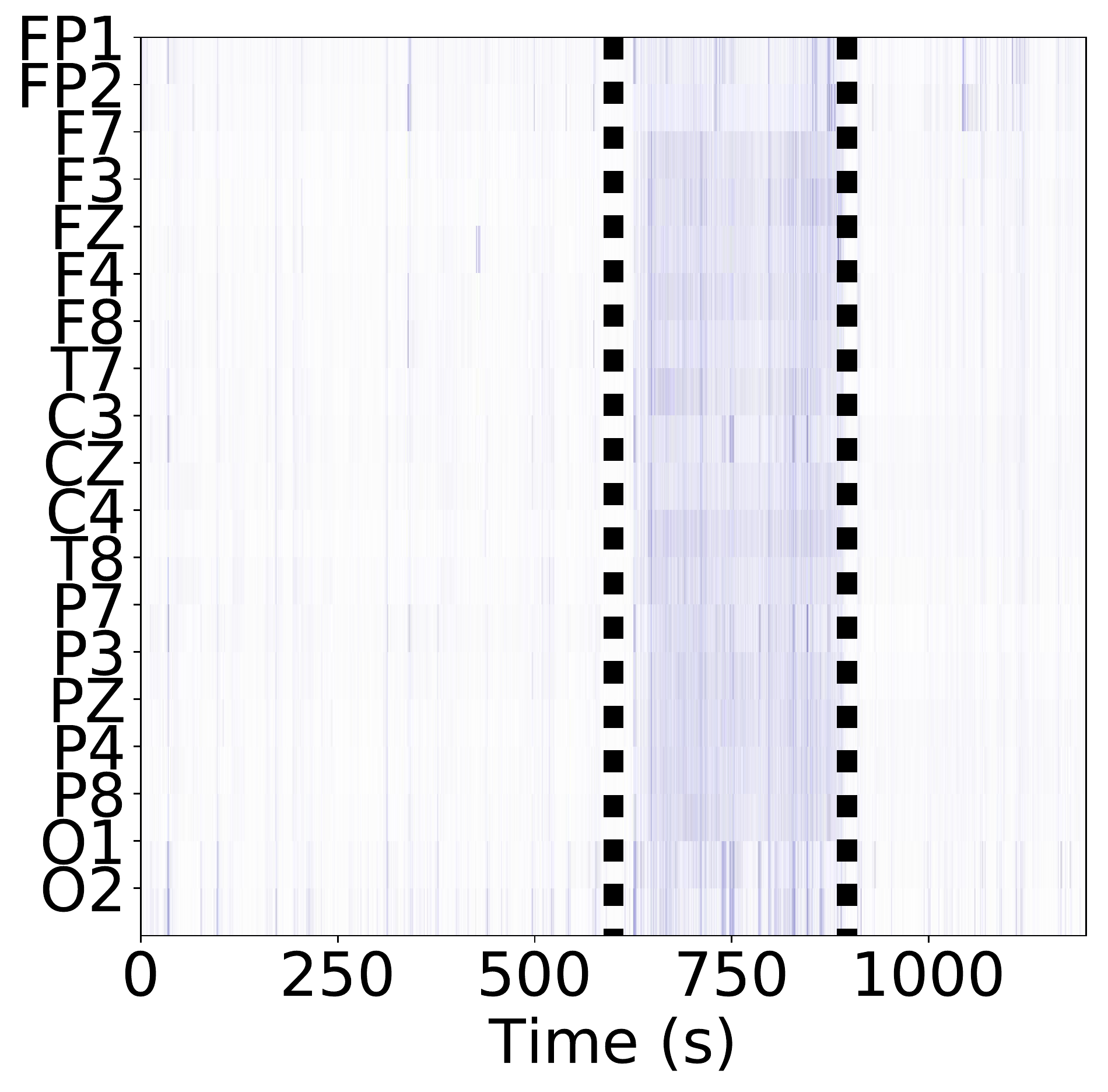}}
  \centerline{\footnotesize (b) Logistic Regression}
\end{minipage}
\hfill
\begin{minipage}[b]{.363\linewidth}
  \centering
  \centerline{\includegraphics[width=\textwidth]{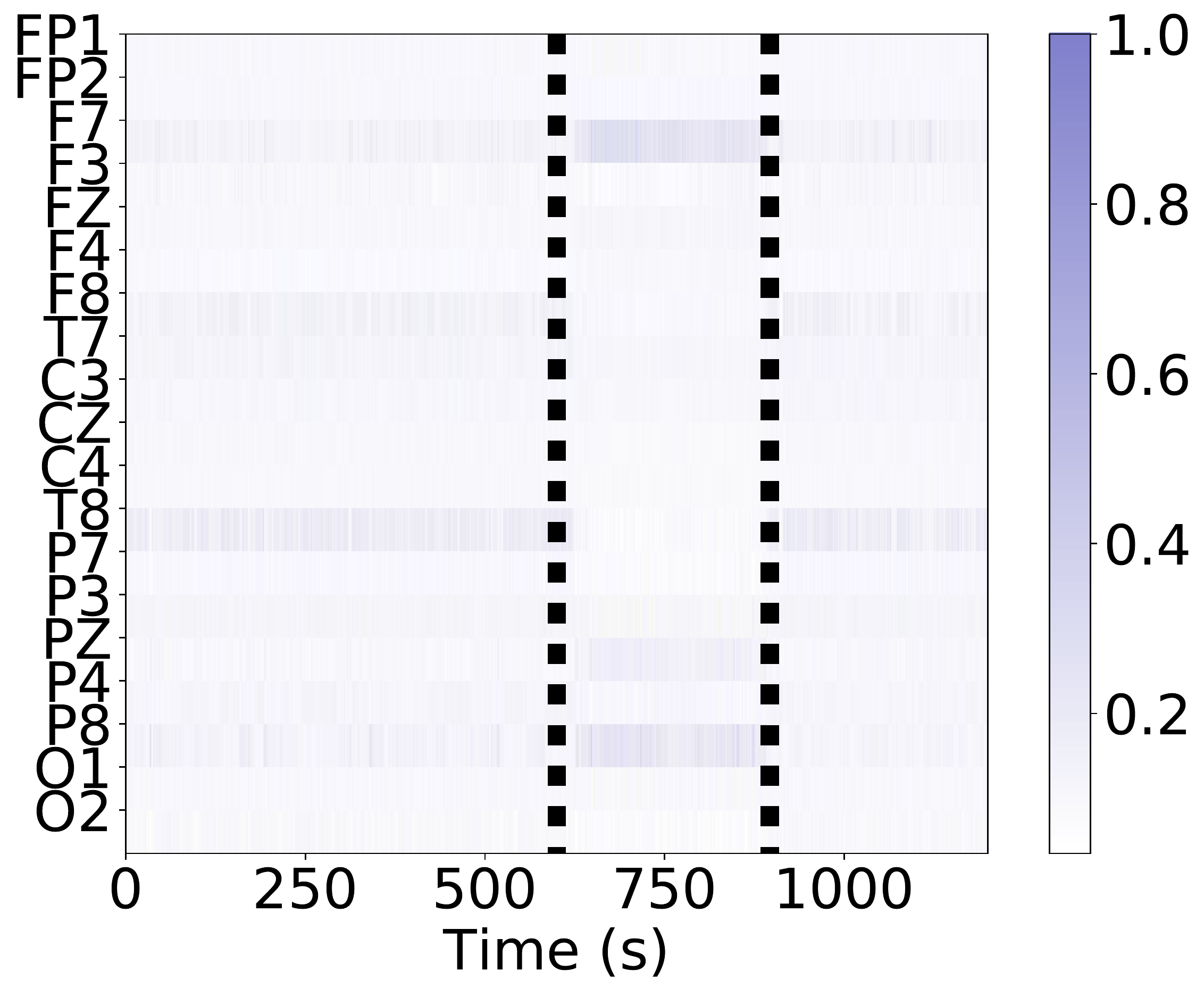}}
  \centerline{\footnotesize (c) SVM}
\end{minipage}

\caption{Detection results on Patient 1 for the three baseline methods. These algorithms place much lower posterior confidence in seizure intervals than the CHMM.}
\label{fig:baselines}
\end{figure}

The transition prior allows our CHMM to place more confidence in contiguous regions exhibiting seizure-like activity. Fig. \ref{fig:inference} (a) shows an example of our classifier correctly classifying the majority of the seizure across all channels. However, this confidence comes with a reduction in specificity, as the classifier tends to associate post-seizure spectral artifacts with seizure as shown in Fig.~\ref{fig:inference}~(b). In future work we will investigate feature selection methods to combat this issue. Fig.~\ref{fig:inference}~(c) shows the evolution of a focal right temporal seizure, which indicates our model's potential to localize epileptic activity on the scalp. This localization is highly relevant to clinical management of epilepsy.

\newlength{\oldintextsep}
\setlength{\oldintextsep}{\intextsep}

\setlength\intextsep{0pt}

\begin{wraptable}[7]{r}{7cm}
\centering
\caption{Results for each method.}
\begin{tabular}{|l|l|l|c|}
	\hline
	Model & TPR & TNR & AUC \\
	\hline
	GMM 					& 21.91\%	& 92.39\%	& 0.784 \\
	Logistic Regression 	& 18.07\%	& \textbf{92.55\%}	& 0.80 \\
	Kernel SVM			& 10.22\%	& 90.27\%	& 0.53 \\
	CHMM 				& \textbf{72.63\%}	& 79.27\% 	& \textbf{0.839} \\
	\hline
\end{tabular}
\label{tab:results}
\end{wraptable}
Due to the heterogeneity of seizure presentations across patients, our baselines fail to perform well as shown in Fig.~\ref{fig:baselines}. The logistic regression and GMM correctly classify portions of seizure intervals but lack consistency. The GMM exhibits more confidence in classifying seizures than its probabilistic linear counterpart. The SVM performs poorly due to the inseperability of the EEG features in our noisy clinical dataset.

\section{Conclusion}
We have presented a novel method for epileptic seizure detection based on a CHMM model. At a high level, we directly model seizure spreading by allowing the state of neighboring and symmetric EEG channels to influence the transition probabilities for a given channel. We have validated our approach on clinical EEG data from 15 unique patients. Our model outperforms three baseline approaches which perform classification on a framewise basis. By incorporating a transition prior that includes spatial and temporal contiguity to seizure regions we are able to better classify seizure intervals within EEG recordings. 

\subsubsection{Acknowledgments:} This work was supported by a Johns Hopkins Medical Institute Synergy Award (Joint PI: Venkataraman/Johnson).

\bibliographystyle{splncs}
\bibliography{miccai_references}

\end{document}